\documentclass[aps,pra,twocolumn,showpacs,preprintnumbers,amsmath,superscriptaddress,amssymb,floats,nofootinbib]{revtex4}

\usepackage{graphicx}% Include figure files
\usepackage{dcolumn}% Align table columns on decimal point

\begin{document}

\def\NIM{\rm Nucl. Instr. Meth.}
\def\NIMA{{\rm Nucl. Instr. and Meth.} A} 
\def\NPB{{\rm Nucl. Phys.} B} 
\def\PLB{{\rm Phys. Lett.}  B} 
\def\PRL{\rm Phys. Rev. Lett.}
\def\PL{\rm Phys. Lett.} 
\def\PRD{{\rm Phys. Rev.} D} 
\def\PRA{{\rm Phys. Rev.} A}
\def\PRC{{\rm Phys. Rev.} C} 
\def\PR{\rm Phys. Rev.}
\def\NPB{{\rm Nucl. Phys.} B}
\def\MRI{\rm J. Magn. Res. Imag.}
\def\MRM{\rm Magn. Res. Med.}
\def\PRT{\rm Phys. Rep.}
\def\JAP{\rm J. Appl. Phys.}
\def\JCP{\rm J. Chem. Phys.}
\def\JMR{\rm J. Magn. Res.}
\def\JMRB{\rm J. Magn. Res. B}
\def\JPB{\rm J. Phys. B}
\def\CPL{\rm Chem. Phys. Lett.}
\def\RMP{\rm Rev. Mod. Phys}
\def\Journal#1#2#3#4{{#1} {#2} (#4) {#3}}

\title{General Solution to Gradient Induced Transverse and Longitudinal Relaxation of Spins Undergoing Restricted Diffusion}
\author{W. Zheng, H. Gao, J.-G. Liu, Y. Zhang, Q. Ye}
\affiliation{Triangle Universities Nuclear Laboratory and Department of Physics, Duke University, Durham, NC 27708, USA}
\author{C. Swank}
\affiliation{Department of Physics, North Carolina State University, Raleigh, NC 27695, USA}

\begin{abstract}

We develop an approach, by calculating the autocorrelation function of spins, to derive the magnetic field gradient induced transverse ($T_2$) relaxation of spins undergoing restricted diffusion. This approach is an extension to the method adopted by McGregor. McGregor's approach solves the problem only in the fast diffusion limit; however, our approach yields a single analytical solution suitable in all diffusion regimes, including the intermediate regime. This establishes a direct connection between the well-known Torrey's slow diffusion result and the fast diffusion result. We also perform free induction decay measurements on spin-exchange optically polarized $^3$He gas with different diffusion constants. The transverse relaxation profiles are compared with the theory and satisfactory agreement has been found throughout all diffusion regimes. In addition to the transverse relaxation, this approach is also applicable to solving the longitudinal relaxation ($T_1$) regardless of the diffusion limits. It turns out that the longitudinal relaxation in the slow diffusion limit differs by a factor of two, compared with that in the fast diffusion limit.

\end{abstract}


\maketitle

\section{Introduction}

Longitudinal relaxation (T$_1$ relaxation) and transverse relaxation (T$_2$ relaxation) are the most important parameters in Nuclear Magnetic Resonance (NMR), Magnetic Resonance Spectroscopy and Magnetic Resonance Imaging (MRI). Many factors can contribute to T$_1$ and T$_2$ relaxations. It is well known that, for liquid or gaseous samples, such as spin polarized $^3$He, diffusion in a non-uniform magnetic field can cause both T$_1$ and T$_2$ relaxations. In this manuscript, $1/T_1^G$ and $1/T_2^G$ represent the relaxation rates solely due to diffusion in a magnetic field gradient. The transverse relaxation due to diffusion in free space was first solved by Torrey \cite{Torrey}. He generalized the Bloch equation by adding a diffusion term and unveiled that the envelope of the transverse component of the magnetization decays as
\begin{equation}
	A(t)=\exp(-\frac{1}{3}D\gamma^2G^2t^3),
	\label{eq:torrey}
\end{equation}
where $D$ is the diffusion constant, $\gamma$ is the gyromagnetic ratio of the spin and $G$ is a constant gradient. In practice, most diffusion happens in confined spaces, and the stochastic diffusion process is restricted, which makes the problem more complicated. Robertson \cite{Robertson} solved the Bloch-Torrey equation \cite{Torrey} in restricted geometries, by imposing boundary conditions to the equation. His approximate analytical solution showed that the envelop decays exponentially with a constant relaxation rate. This result is valid when $4Dt>>L^2$, where L is the distance between the boundaries. This limit is also known as fast diffusion limit or motional averaging regime, where spins have moved across the geometry many times in a time period $t$ and therefore any fluctuation in the magnetic field averages out and a faster diffusion actually reduces the relaxation. Neuman \cite{Neuman} solved the same problem by calculating the accumulated phases of spins with the assumption that the relative phase distribution of spins is Gaussian in both the slow diffusion and fast diffusion limits. His slow diffusion result reproduces the free diffusion result, Eq. (\ref{eq:torrey}), and the fast diffusion result is the same as that of Robertson. In the intermediate regime, the Gaussian Phase Approximation (GPA) fails. However, it is crucial to quantitatively understand the intermediate regime because many experiments have shown edge enhancement phenomena in the slow diffusion and intermediate regime, which was recognized later as a localization regime \cite{Hayden}. People have observed that, when water diffuses in microscopic structures, the MRI signal is enhanced at the edge of the structure \cite{Putz,Hyslop,Barsky,CallaghanPaper,Hurlimann}. Saam \textit{et al.} have also showed a similar edge enhancement effect, using hyperpolarized $^3$He gas in cells with dimensions of about 1 cm \cite{Saam}. This effect is ascribed to the more restricted diffusion at the boundary, which lessens the relaxation, and was first described quantitatively by De Swiet \cite{Swiet}, using Airy functions. Airy functions have been shown to be the eigenfunction of the Bloch-Torrey equation in the intermediate regime \cite{Stoller}. Axelrod also showed that although GPA fails in the intermediate regime, it can be used to interpolate the result in this regime, which turns out to be close to the exact solution \cite{Axelrod}. More detailed discussion on the restricted diffusion in various limits can be found in a review article \cite{Grebenkov} and references therein.

Despite of the widely used GPA method, Cates \textit{et al.} used the second order time-dependent perturbation theory and carried out an expansion of spin density matrix to obtain both longitudinal and transverse relaxation rates for a spherical cell \cite{Cates}. Their results works only in the fast diffusion regime, and they further divided the fast diffusion regime into two limits: the high pressure limit $\omega_0R^2/8\pi D>>1$ and the low pressure limit $\omega_0R^2/8\pi D<<1$, where $\omega_0$ is the Larmor precession frequency and $R$ is the radius of the spherical cell. These two limits can be thought of as the characteristic spin precession time $\tau_l=2\pi/\omega_0$ being much shorter or longer than the characteristic diffusion time in the cell $\tau_d=R^2/4D$, respectively. McGregor \cite{McGregor} used Redfield theory discussed in \cite{Slichter} to solve the same problem also in the fast diffusion regime. Redfield theory is a generalized treatment of the second-order time-dependent perturbation theory. It establishes a set of differential equations obeyed by the spin density matrix. Therefore, it is closely related to the treatment of Cates \textit{et al.} \cite{Golub}. By calculating autocorrelation function of spin under different geometries, McGregor was able to obtain transverse relaxation rates in the fast diffusion limit for different geometries, including slabs, cylinders and spheres.

In this manuscript, we make an extension to McGregor's approach, which yields an analytical solution to the transverse magnetization suitable for all diffusion regimes. When $4Dt<<L^2$, this solution reproduces Torrey's free diffusion result; when time $4Dt\approx L^2$, it is in the intermediate regime. In these two regimes, edge enhancement effect is also observed. Eventually, when $4Dt>>L^2$, it is in the motional averaging regime and a peak located at the center of the frequency spectrum is observed. We also performed Free Induction Decay (FID) measurements on polarized $^3$He gas to verify the theoretical results. By changing the number density of the gas, the observed transverse relaxation happens in different diffusion regimes. When the decay envelopes are compared to the theoretical predictions, they are found to be in good agreements, especially in the intermediate regime. In addition to the transverse relaxation, our approach can also be used to calculate longitudinal relaxation in different regimes. We found that the longitudinal relaxation rate $1/T_1^G$ in the slow diffusion limit is twice as fast as that in the fast diffusion limit. As diffusion in the fast diffusion regime is more restricted, it could explain this factor of two difference. In this manuscript, we solve the problem in 1D for clarification purpose. However, it can be easily extended to 3D with complex geometries since one only needs to calculate the corresponding probability density function. Once the density function is known, the relaxation rate can be calculated readily, which makes this approach suitable for numerical simulations of complex geometries.

\section{Redfield theory for magnetic field gradient-induced relaxations}

For simplicity, let spins diffuse in a cubic cell with length $L$. A non-uniform magnetic field is applied along the $\hat{z}$ direction. At time $t=0$, we track a spin starting at $\vec{x}^{\prime}_c$. As time evolves, the expected position of the spin will change due to the diffusion process. Therefore, we use $\left\langle \vec{x}^{\prime}(t)\right\rangle$ to represent the expected position of spin at some later time $t$ and has the property that $\left\langle \vec{x}^{\prime}(0)\right\rangle=\vec{x}^{\prime}_c$. Since the field is non-uniform over the box, the spins also see fluctuating magnetic fields during diffusion. The fluctuating field $\vec{B}^{\prime}$ can be treated as a perturbation to the zeroth order mean field $\vec{B}(\left\langle \vec{x}^{\prime}\right\rangle)$ by taking Taylor expansion around $\vec{x}=\left\langle \vec{x}^{\prime}\right\rangle$,
\begin{equation}
	\vec{B}^{\prime}\equiv \vec{B}(\vec{x}(t))-\vec{B}(\left\langle \vec{x}^{\prime}(t)\right\rangle)=\vec{\nabla} \vec{B}\cdot(\vec{x}(t)-\left\langle \vec{x}^{\prime}(t)\right\rangle).
	\label{eq:xyz}
\end{equation}
As described in \cite{Slichter}, Redfield theory gives solutions to the problem with fluctuating magnetic fields. In our case, the applied field is constant in time; however the time dependence appears because of the diffusion process. The equation of motion for the transverse and longitudinal components of the spin can be written as \cite{Slichter}
\begin{align}
	\frac{d}{dt}\left\langle S_{T}\right\rangle &\equiv\frac{d}{dt}\left(\left\langle S_{x}\right\rangle+i\left\langle S_{y}\right\rangle\right)=\gamma (\left\langle \textbf{S}\textbf{$_T$}\right\rangle\times \textbf{B})- \notag \\ 
	&\gamma ^{2}\left\{\frac{1}{2}\left[J_{B^{\prime}_{x}}(\omega)+J_{B^{\prime}_{y}}(\omega)\right]+J_{B^{\prime}_z}(0)\right\}\left\langle S_{T}\right\rangle
	\label{eq:t2eqn}
\end{align}

\begin{align}
	\frac{d}{dt}\left\langle S_{z}\right\rangle &=-\gamma ^{2}\left[J_{B^{\prime}_{x}}(\omega)+J_{B^{\prime}_{y}}(\omega)\right]\left\langle S_{z}\right\rangle,
	\label{eq:t1eqn}
\end{align}
where $S_T$ is the transverse component of spins, $\textbf{B}=B_z(\left\langle \vec{x}^{\prime}\right\rangle)\hat{z}$, $\omega=\gamma B_z(\left\langle \vec{x}^{\prime}\right\rangle)$, the Larmor precession frequency of spins at field strength $B_z(\left\langle \vec{x}^{\prime} \right\rangle)$, and $J_{B^{\prime}_x}$ is defined as \cite{Slichter}
\begin{equation}
	J_{B^{\prime}_{x}}(\omega)=\int^{t}_{0}\overline{B^{\prime}_{x}(t-\tau)B^{\prime}_{x}(t)}e^{-i\omega\tau}d\tau.
	\label{eq:bautocorr}
\end{equation}
The bar denotes an ensemble average of the autocorrelation of the perturbed magnetic field. $J_{B^{\prime}_z}(0)$ has the similar definition with $B^{\prime}_x$ replaced by $B^{\prime}_z$ and $\omega=0$. The first term in Eq. (\ref{eq:t2eqn}) describes the precession of the spin under the field $B_z(\left\langle \vec{x}^{\prime} \right\rangle)$, and the second term gives the transverse relaxation rate $1/T_2^G$; whereas Eq. (\ref{eq:t1eqn}) describes the longitudinal relaxation. 

It should be noted that Eq. (\ref{eq:bautocorr}) was originally written in \cite{Slichter} as an integral from 0 to infinity. It was argued that the autocorrelation of the magnetic field would vanish quickly after a critical time $\tau_c$, and consequently, integration from 0 to infinity introduces negligible errors as long as $t>>\tau_c$. For the case considered here, $\tau_c$ can be defined as $\tau_c\equiv (L/2)^2/4D$. When the diffusion is slow, the above approximation is invalid, and Eq. (\ref{eq:bautocorr}) must be used. By utilizing Eq. (\ref{eq:xyz}) and realizing the fact that diffusions in $\hat{x}$, $\hat{y}$ and $\hat{z}$ directions are independent of each other, $J_{B^{\prime}_{x}}(\omega)$ can be re-written as
\begin{equation}
	J_{B^{\prime}_{x}}(\omega)=(\frac{\partial B^{\prime}_x}{\partial x})^{2}J_{x}(\omega)+(\frac{\partial B^{\prime}_x}{\partial y})^{2}J_{y}(\omega)+(\frac{\partial B^{\prime}_x}{\partial z})^{2}J_{z}(\omega),
	\label{eq:xautofourier}
\end{equation}
where $J_x(\omega)$ is expressed as
\begin{equation}
	J_x(\omega)=\int^{t}_{0}\overline{(x(t-\tau)- \left\langle x^{\prime}(t)\right\rangle)(x(t)- \left\langle x^{\prime}(t)\right\rangle)}e^{-i\omega\tau}d\tau,
	\label{eq:Jxexp}
\end{equation}
and similar for $J_y(\omega)$ and $J_z(\omega)$. In general, $J_x(\omega)$ is time-dependent and the relaxation rates are also time-dependent. However, it will be shown that, in the fast diffusion limit, $J_x(\omega)$ approaches some constant values and $J_x(\omega)=J_y(\omega)=J_z(\omega)$. Hence, one can define the longitudinal relaxation rate $1/T_1^G$ as
\begin{equation}
	\frac{1}{T_1^G}=\gamma ^{2}(\left|\nabla B_x\right|^2+\left|\nabla B_y\right|^2)\Re\left[J_{x}(\omega)\right],
	\label{eq:t1expression}
\end{equation}
where $\Re\left[\cdot\right]$ indicates taking the real part. Hence, in the fast diffusion limit, the transverse relaxation rate $1/T_2^G$ can also be defined as
\begin{equation}
	\frac{1}{T_2^G}=\frac{1}{2T_1^G}+\gamma^2\left|\nabla B_z\right|^2J_{x}(0).
	\label{eq:t2expression}
\end{equation}

\subsection{Magnetic Field Gradient-induced Transverse Relaxation Rate $1/T_2^G$}

The expected position of spin $\left\langle x^{\prime}(t)\right\rangle$ at time $t$ can be computed as
\begin{equation}
	\left\langle x^{\prime}(t)\right\rangle=\int^{L/2}_{-L/2}x\rho(x,t|x^{\prime}_c,0)dx,
	\label{eq:xavg}
\end{equation}
where $\rho(x,t|x_{0},t_{0})$ is the conditional probability density function of spin at position $x$ at time $t$, if the spin is at $x_{0}$ at time $t_{0}$ ($t_0=t-\tau$). It is apparently that Eq. (\ref{eq:xavg}) satisfies the condition $\left\langle x^{\prime}(0)\right\rangle=x^{\prime}_c$.

In order to compute $J_x(0)$, one needs to calculate position autocorrelation function, which, by definition, can be expressed as
\begin{widetext}
\begin{equation}
	\overline{(x(t-\tau)- \left\langle x^{\prime}\right\rangle)(x(t)- \left\langle x^{\prime}\right\rangle)}=\int^{L/2}_{-L/2}\int^{L/2}_{-L/2} (x_{0}-\left\langle x^{\prime}\right\rangle)(x-\left\langle x^{\prime}\right\rangle)\rho(x,t|x_{0},t_{0})\rho(x_{0},t_{0}|\left\langle x^{\prime}\right\rangle,0)dxdx_{0}.
	\label{eq:xauto}
\end{equation}
\end{widetext}
Due to the symmetry of the cubic cell, $\rho(x,t|x_{0},t_{0})$ can be found by decomposing the 3D diffusion equation into 1D problem, subject to the boundary condition 
	\[\frac{\partial}{\partial x}\rho(x,t|x_0,t_0)|_{x=\pm L/2}=0,
\]
and the initial condition
	\[\rho(x,t_{0}|x_{0},t_{0})=\delta(x-x_{0}).
\]
The solution is found to be

\begin{align}
		\rho(x,t|x_0,t_0) &=\frac{1}{L}+\frac{2}{L}\sum^{\infty}_{n=1,3}e^{-\frac{n^2\pi^2D\tau}{L^2}}\sin(\frac{n\pi x}{L})\sin(\frac{n\pi x_0}{L}) \notag \\ &+\frac{2}{L}\sum^{\infty}_{n=2,4}e^{-\frac{n^2\pi^2D\tau}{L^2}}\cos(\frac{n\pi x}{L})\cos(\frac{n\pi x_0}{L}).
	\label{eq:gensolution}
\end{align}

By substituting Eq. (\ref{eq:gensolution}) into Eq. (\ref{eq:xauto}), one obtains
\begin{widetext}
\begin{align}
	&\overline{(x(t-\tau)- \left\langle x^{\prime}\right\rangle)(x(t)- \left\langle x^{\prime}\right\rangle)}=\frac{8L^2}{\pi^4}\sum^{\infty}_{n=1,3}\frac{1}{n^4}\exp(-\frac{n^2\pi^2D\tau}{L^2})+ \left\langle x^{\prime}\right\rangle^2\notag\\ 
	&-\frac{4\left\langle x^{\prime}\right\rangle L}{\pi^2}\sum^{\infty}_{n=1,3}\frac{1}{n^2}\exp(-\frac{n^2\pi^2Dt}{L^2})\sin(\frac{n\pi \left\langle x^{\prime}\right\rangle}{L})\sin(\frac{n\pi}{2}
)\notag\\
&-\frac{4 \left\langle x^{\prime}\right\rangle L}{\pi^2}\sum^{\infty}_{k=2,4}\frac{1}{k^2}\exp(-\frac{k^2\pi^2Dt_0}{L^2})\sin(\frac{k\pi \left\langle x^{\prime}\right\rangle}{L})\sin(\frac{k\pi}{2})\notag\\
&+\frac{16L^2}{\pi^4}\sum^{\infty}_{n=1,3}\sum^{\infty}_{k=2,4}\frac{k^2+n^2}{n^2(k^2-n^2)^2}\exp(-\frac{n^2\pi^2D\tau}{L^2})\exp(-\frac{k^2\pi^2Dt_0}{L^2})\cos(\frac{k\pi  \left\langle x^{\prime}\right\rangle}{L})\cos(\frac{k\pi}{2}).
\label{eq:xautogensolution}
\end{align}
\end{widetext}
Instead of substituting Eq. (\ref{eq:xautogensolution}) into $J_x$ directly, one can first take the slow diffusion or fast diffusion limits on it. It can be shown (see Appendix) that, in the slow diffusion limit $4Dt<<(L/2)^2$, Eq. (\ref{eq:xautogensolution}) is simplified to
\begin{equation}	
\overline{(x(t-\tau)-\left\langle x^{\prime}\right\rangle )(x(t)- \left\langle x^{\prime}\right\rangle)}\approx2D(t-\tau).
\label{eq:freeautocorr}
\end{equation}
$J_x(0)$ is then calculated as
\begin{equation}
	J_x(0)=\int^{t}_{0}2D(t-\tau)d\tau=Dt^2.
	\label{eq:slowJx}
\end{equation}
This is also true for $J_y(0)$ and $J_z(0)$. As $J_{x,y,z}(0)$ is time-dependent, the relaxation rate is also time-dependent. One has to substitute $J_{x,y,z}(0)$ into Eq. (\ref{eq:t2eqn}) directly to compute $S_T$,
\begin{equation}
	\frac{d}{dt}\left\langle S_{T}\right\rangle=\gamma (\left\langle \textbf{S$_T$}\right\rangle\times \textbf{B}(\left\langle \vec{x}^{\prime}\right\rangle))-(\frac{1}{2T_1}+\gamma^2|\nabla B_z|^2Dt^2)\left\langle S_{T}\right\rangle.
	\label{eq:eqofmoslow}
\end{equation}

If one assumes that the gradient in one direction $\partial B_z/\partial x \equiv G$ dominates, then $|\nabla B_z|\approx G$ and $B_z=B_z(0)+G\left\langle x^{\prime}\right\rangle\approx B_z(0)+G x^{\prime}_c$. $\left\langle x^{\prime}\right\rangle\approx x^{\prime}_c$ because spins are localized in the slow diffusion limit and the expected position at time $t$ will not change significantly from their initial positions. Integrating upon time, Eq. (\ref{eq:eqofmoslow}) becomes
\begin{equation}
	\left\langle S_{T}\right\rangle=S_0\exp[i\gamma(B_z(0)+G x^{\prime}_c)t]\exp(-\frac{\gamma^2G^2Dt^3}{3}+\frac{t}{2T_1^G}).
	\label{eq:slowSx}
\end{equation}
This coincides with 1D diffusion result derived by Torrey \cite{Torrey}. Torrey's derivation assumes free diffusion without boundaries, which is equivalent to the slow diffusion in a confined volume, as boundaries are not present to spins when diffusion is slow.

In the fast diffusion limit, $4Dt>>(L/2)^2$, all the exponential terms containing $t$ in Eq. (\ref{eq:xautogensolution}) vanish, so does $\left\langle x^{\prime}\right\rangle^{2}$ term, which can be proved easily by taking the limit $4Dt/L^2\rightarrow\infty$ in Eq. (\ref{eq:xavg}). Therefore, the only surviving term is the first one,
\begin{equation}
\overline{(x(t-\tau)- \left\langle x^{\prime}\right\rangle)(x(t)-\left\langle x^{\prime}\right\rangle)}=\sum^{\infty}_{n=1,3}\frac{8L^2}{n^4\pi^4}e^{-\frac{n^2\pi^2D\tau}{L^2}}.
	\label{eq:fastdiff}
\end{equation}
$J_x(0)$ in this case becomes
\begin{equation}
	J_x(0)=\int^{t}_{0}\sum_{n=1,3}\frac{8L^2}{n^4\pi^4}e^{-\frac{n^2\pi^2D\tau}{L^2}}d\tau=\frac{L^4}{120D},
	\label{eq:fastJx}
\end{equation}
where the fact that $4Dt>>(L/2)^2$ is used again after the integration. $J_y(0)$ and $J_z(0)$ are the same as $J_x(0)$ in this limit. Substituting them back into Eq. (\ref{eq:t2eqn}), one obtains
\begin{equation}
	\left\langle S_{T}\right\rangle=S_0\exp[i\omega_0t-(\frac{1}{2T_1}+\frac{\gamma^2L^4\left|\nabla B_z\right|^2}{120D})t].
	\label{eq:fastSx}
\end{equation}
As $\left\langle x^{\prime}(t)\right\rangle$ approaches zero in the fast diffusion limit, all the spins precess at the same central frequency $\omega_0=\gamma B_z(0)$, no matter where the spin is initially. This is also known as motional averaging regime. If gradient in one direction dominates the other two ($\left|\nabla B_z\right|\approx G$), the transverse component decays with a constant relaxation rate $1/T_2^G$ given by 
\begin{equation}
	\frac{1}{T_2^G}=\frac{1}{2T_1^G}+\frac{\gamma^2L^4G^2}{120D}.
	\label{eq:fastT2}
\end{equation}
This result is also derived by McGregor \cite{McGregor} and Robertson \cite{Robertson} using GPA method. However, it will be shown in the discussion section that, by numerically calculating $J_x(0)$ using Eq. (\ref{eq:xautogensolution}) without any approximation, one can obtain the frequency spectrum of the precession signal in various limits, which cannot be obtained from McGregor and Robertson's methods. As a constant gradient is applied, the frequency spectrum is actually a frequency encoded 1D image. In the slow diffusion limit, peaks are observed at the edge of the geometry, known as edge enhancement; whereas in the fast diffusion limit, a resonance peak is observed at the center of the spectrum.

In the intermediate region $4Dt\approx(L/2)^2$, one has to substitute Eq. (\ref{eq:xautogensolution}) directly into $J_x(0)$ and then calculate Eq. (\ref{eq:t2eqn}). Unfortunately, no concise analytical form of $S_T(\left\langle x^{\prime}\right\rangle)$ can be obtained in this regime. Hence, a numerical calculation of $S_T(\left\langle x^{\prime}\right\rangle)$ is performed and compared with Free Induction Decay (FID) measurements on gaseous $^3$He cells, and a good agreement in the intermediate regime is shown, see Sec. III.

\subsection{Magnetic Field Gradient-induced Longitudinal Relaxation Rate $1/T_1^G$}
{
In order to compute Eq. (\ref{eq:t1eqn}), one needs to obtain $J_x(\omega)$ first. In the slow diffusion limit ($4Dt<<(L/2)^2$), Eq. (\ref{eq:freeautocorr}) should be used,
\begin{align}
	J_{x}(\omega)&=\int^{t}_{0}2D(t-\tau)e^{-i\omega\tau}d\tau\notag\\
	&=\frac{2D(1-e^{-i\omega t}-i\omega t)}{\omega^2}.
	\label{eq:JBxnonstat}
\end{align}
Since, in most cases, $\omega>>1$, $\Re(1-e^{-i\omega t})=1-\cos\omega t$ is a fast oscillating function, which averages to 1. Consequently, substituting $\Re[J_x(\omega)]$ into Eq. (\ref{eq:t1expression}), the longitudinal relaxation has an averaged decay rate as
\begin{equation}
\frac{1}{T_1^G}=\gamma^2(\left|\nabla B_x\right|^2+\left|\nabla B_y\right|^2)\frac{2D}{\omega^2}=2D\frac{\left|\nabla B_x\right|^2+\left|\nabla B_y\right|^2}{B_0^2},
\label{eq:nonstatT1}
\end{equation}
where $\omega=\gamma B_z(\left\langle \vec{x}^{\prime}\right\rangle)\approx\gamma B_z(0)\equiv\gamma B_0$.

In the fast diffusion limit ($4Dt>>(L/2)^2$), Eq. (\ref{eq:fastdiff}) should be used to calculate $J_{x}(\omega)$, 
\begin{align} 
	\label{eq:Jxstationary}
	J_{x}(\omega) &=\frac{8L^2}{\pi^4}\int^{\infty}_{0}\sum^{\infty}_{n=1,3}\frac{1}{n^4}e^{-\frac{n^2\pi^2D\tau}{L^2}}e^{-i\omega\tau}d\tau \notag \\
	&= \frac{8L^2}{\pi^4}\sum^{\infty}_{n=1,3}\frac{1}{n^4}\frac{1}{\frac{n^2\pi^2D}{L^2}+i\omega}.
\end{align}
Hence, the real part of $J_{x}(\omega)$ is written as
\begin{equation}
	\Re[J_{x}(\omega)]=\frac{8L^4D}{\pi^2}\sum^{\infty}_{n=1,3}\frac{1}{n^2}\frac{1}{n^4\pi^4D^2+\omega^2L^4}.
	\label{eq:ReJxstat}
\end{equation}

In the fast diffusion limit with high pressure ($\tau_d/\tau_l=\omega L^2/32\pi D>>1$), the sum in Eq. (\ref{eq:ReJxstat}) is simplified to
\begin{equation}
	\sum^{\infty}_{n=1,3}\frac{1}{n^2}\frac{1}{\omega^2L^4}=\frac{\pi^2}{8\omega^2L^4},
\end{equation}
and $1/T_1^G$ in this limit becomes
\begin{equation}
	\frac{1}{T_1^G}=D\frac{\left|\nabla B_x\right|^2+\left|\nabla B_y\right|^2}{B_0^2},
	\label{eq:T1eqn}
\end{equation}
It is interesting to see that there is a factor of 2 difference between the slow diffusion $1/T_1^G$ and the high pressure fast diffusion $1/T_1^G$. More discussion on this topic is presented in Sec. IV. 

In the fast diffusion with low pressure ($\tau_d/\tau_l=\omega L^2/32\pi D<<1$), the sum in Eq. (\ref{eq:ReJxstat}) becomes
\begin{equation}
	\sum^{\infty}_{n=1,3}\frac{1}{n^6}\frac{1}{\pi^4D^2}=\frac{\pi^2}{960D^2},
\end{equation}
and the resultant $1/T_1^G$ in the low pressure limit is
\begin{equation}
	\frac{1}{T_1^G}=\frac{\gamma^2L^4}{120D}(\left|\nabla B_x\right|^2+\left|\nabla B_y\right|^2).
	\label{eq:T1eqnlp}
\end{equation}
This result is an analogy to the low pressure $1/T_1^G$ derived by Cates \textit{et al.} for a spherical cell geometry \cite{Cates}. For geometries other than sphere and box, one only needs to recalculate Eq. (\ref{eq:xautogensolution}) and the corresponding $1/T_1^G$ can be obtained readily through steps illustrated above. This also applies to the transverse relaxation rate $1/T_2^G$ when other geometries are considered.

\subsection{Magnetic Field Gradient-induced Resonance Frequency Shift}
In Eq. (\ref{eq:t2eqn}), the imaginary part of the complex function $J_{B_x^{\prime}}$ and $J_{B_y^{\prime}}$ gives rise to the shift of precession frequency $\delta\omega$, 
\begin{equation}
	\delta\omega=\frac{-\gamma^2}{2}\left[\left|\nabla B_x\right|^2+\left|\nabla B_y\right|^2\right]\Im\left[J_x(\omega)\right],
	\label{eq:freqshift}
\end{equation}
where $\Im\left[\cdot\right]$ means taking the imaginary part. In the slow diffusion limit, substituting Eq. (\ref{eq:JBxnonstat}) into Eq. (\ref{eq:freqshift}) yields
\begin{equation}
		\delta\omega=\frac{\gamma Dt}{B_0}(\left|\nabla B_x\right|^2+\left|\nabla B_y\right|^2).
		\label{eq:nonstatshift}
\end{equation}
It is interesting to note that, in the slow diffusion limit, the frequency shift increases linearly as a function of time, different from the $t^3$ dependence in the transverse relaxation rate. In addition,  Eq. (\ref{eq:nonstatshift}) does not depend on $L$ as expected because the slow diffusion limit is equivalent to the free diffusion, in which spins do not see boundaries.

In the fast diffusion limit with high pressures, substituting Eq. (\ref{eq:Jxstationary}) into Eq. (\ref{eq:freqshift}) and taking the corresponding limit yields
\begin{equation}
	\delta\omega=\frac{\gamma^2L^2}{12\omega_0}(\left|\nabla B_x\right|^2+\left|\nabla B_y\right|^2)
	\label{eq:statfreqshift}
\end{equation}

In the fast diffusion limit with low pressures, it yields
\begin{equation}
	\delta\omega=\frac{17\omega_0\gamma^2L^8}{20160D^2}(\left|\nabla B_x\right|^2+\left|\nabla B_y\right|^2)
\end{equation}
These two results are analogies to the frequency shifts derived in \cite{Cates} for a spherical cell.
}

\section{Experiments and Results}

\begin{figure}
	\centering
		\includegraphics[width=0.4\textwidth]{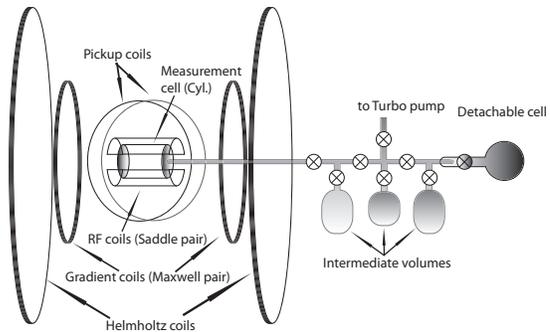}
		\caption{The apparatus for measuring the transverse relaxation of $^3$He in the cylindrical measurement cell. $^3$He in the detachable cell has been polarized by SEOP before it is transferred to the measurement cell.}
	\label{fig:expapp}
\end{figure}

FID measurements have been performed to measure transverse relaxation of polarized $^3$He gas at 34.5 kHz. The $^3$He gas is polarized in a 2 inch diameter spherical detachable cell via Spin-Exchange Optical Pumping (SEOP) technique. The cell is routinely filled with 760 torr $^3$He and 100 torr N$_2$. After $^3$He is polarized, the detachable cell is connected to a measurement cell sitting at the center of the Helmholtz coils through a 2 mm inner diameter glass tube. The measurement cell is made of bare pyrex glass in a cylindrical shape with 2 inch for both the diameter and the length. Three intermediate volumes made of pyrex and a turbo pumping line are connected to the glass transfer tube between the detachable cell and the measurement cell. The schematics of the experimental apparatus is shown in Fig. \ref{fig:expapp}. Before the measurement, polarized $^3$He atoms are allowed to diffuse into intermediate volumes first so that the number density of $^3$He in the detachable cell is diluted. The intermediate volumes are then isolated and the valve between the detachable cell and the measurement cell is opened. Consequently, the rest of $^3$He in the detachable cell can diffuse freely into the measurement cell. When the pressure in the two cells reaches equilibrium, the valve is closed and FID measurements are carried out.

A pair of 6 inch diameter pickup coils in Helmholtz coil configuration is used, so that the coil has a rather uniform sensitivity over the entire cylindrical cell. Each coil has 2000 turns of 30 AWG wires to maximize the signal. The precession signal at 34.5 kHz is lock-in detected and the envelop of the signal is extracted. The RF coil is a pair of saddle coils with a length of 3.5 inch, a diameter of 3 inch and the opening angle is 120 degrees. Each coil has 10 turns of AWG 22 wires. The axis of the pickup coil, the RF field direction and the magnetic holding field are perpendicular to each other. A RF pulse with a tipping angle $\sim20$ degrees is sent to the measurement cell. A pair of gradient coils in Maxwell coil settings \cite{Garrett} is also added to provide a uniform field gradient of 2.3 mG/cm in the holding field or $\vec{z}$ direction. The background gradients are measured to be much smaller than this value and therefore ignored in the calculation shown below.

\begin{figure}
	\centering
		\includegraphics[width=0.4\textwidth]{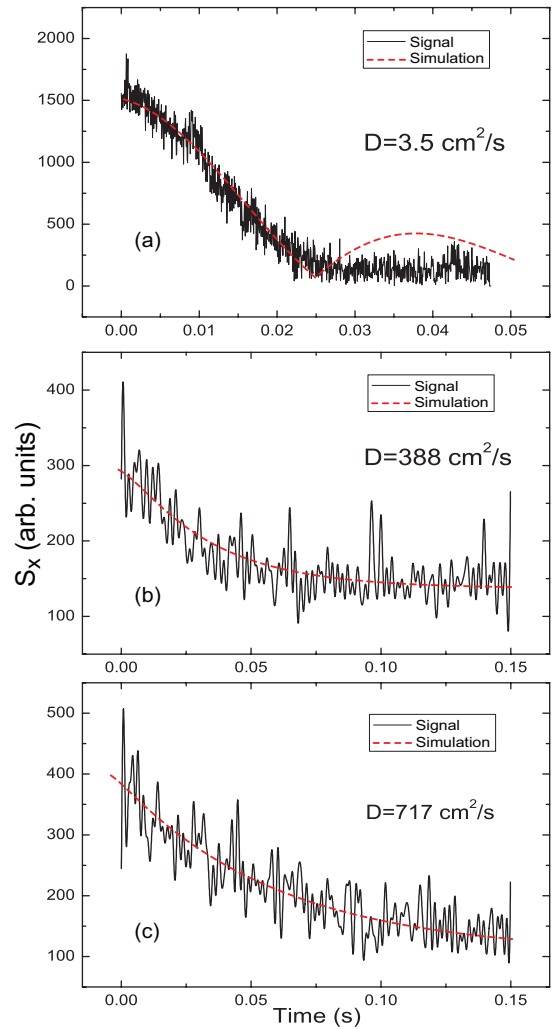}
		\caption{(Color online)Transverse relaxation measurements with different diffusion constants. (a) is in the slow diffusion regime and the sinc like shape of the decay profile is due to the spin defocus, (b) is in the intermediate regime and (c) is in the fast diffusion regime. Simulation results are shown as dashed lines and compared to the measured decay envelops.}
	\label{fig:expT2}
\end{figure}

In Fig. (\ref{fig:expT2}), we show the transverse relaxation measurements of $^3$He with pressures of 327 torr, 2.94 torr and 1.59 torr. The corresponding diffusion constants are 3.5, 388 and 717 cm$^2$/s, respectively (scaled from 1.5 cm$^2$/sec for 760 torr $^3$He, which is measured using Carr-Purcell-Meiboom-Gill method \cite{Callaghan}). When using the ratio $(L/2)^2/4Dt$ to characterize the diffusion regime, it is clear that 3.5 cm$^2$/s is in the slow diffusion regime; whereas 717 cm$^2$/s is in the fast diffusion regime and 388 cm$^2$/s is in the intermediate regime close to fast diffusion limit. When diffusion is slow, the decay of signal is mainly due to the dephasing of the spin precessing at different frequencies. This can be understood by taking $D$ as zero, so that the exponential term in Eq. (\ref{eq:slowSx}) is identical to one and a rough estimate of the overall signal $S_{all}$ is simply integrating $S_T$ over the entire cell,
\begin{equation}
	S_{all}\approx\int^{-L/2}_{L/2}e^{i(\omega_0+\gamma G x^{\prime})}dx^{\prime}\propto e^{i\omega_0t}\text{sinc}(\frac{\gamma GLt}{2})
\end{equation}
where sinc(x) is defined as $\sin(x)/x$ and the first zero of sinc function is at $\gamma GLt=2\pi$. Using the real values of $G$ and $L$, one can determine $t=26$ ms. This value is the same as the experimentally measured signal vanishing time $t\approx26$ ms for the case D=3.5 cm$^2$/s, see Fig. (\ref{fig:expT2}). 

When diffusion is fast, spins will more or less precess at the same frequency (motional averaging) and the diffusion term dedicates the signal decay. As no simple form of $S_T$ can be obtained in the intermediate regime, we numerically calculate $S_T(\left\langle x^{\prime}\right\rangle)$ using the general form of autocorrelation function, Eq. (\ref{eq:xautogensolution}). To evaluate Eq. (\ref{eq:xautogensolution}), we compute one hundred terms in each of the first three sums and four hundred terms in the last double sum. We also evaluate $\left\langle x^{\prime}\right\rangle$ up to 100 terms. Once $S_T(\left\langle x^{\prime}\right\rangle)$ is known, it is weighted by $B(\left\langle x^{\prime}\right\rangle)$, the sensitivity of the Helmholtz pickup coil at position $\left\langle x^{\prime}\right\rangle$, and then integrated over the entire cylindrical cell to mimic the measured FID signal. The simulation results, shown as red dashed curves, are compared with measured decay envelops. The background noise of the FID measurement is around 150 arb. units. The simulation curves are shifted up by this amount to account for the background. Good agreements between measurements and simulations are found for the intermediate regime and the fast diffusion regime. One can also use Eq. (\ref{eq:fastT2}) to predict $T_2$ in Fig. (\ref{fig:expT2}c) as it is in the fast diffusion regime. The prediction yields $T_2=0.0589$ s and an exponential fit of the data yields $T_2=0.0557$ s, which is very close to the prediction. However, if Eq. (\ref{eq:fastT2}) is used to predict $T_2$ in Fig. (\ref{fig:expT2}b), it overestimates $T_2$ by 39$\%$ and the profile of the measured relaxation is somewhere between the sinc and exponential. This shows that Eq. (\ref{eq:fastT2}) is inadequete to use in the intermediate regime, and one has to use the non-approximated form of $S_T(\left\langle x^{\prime}\right\rangle)$ to do the calculation.

In the slow diffusion regime, the numerical calculation correctly captures the time when signal vanishes, and it also exhibits a small bump at 0.04 s, due to the partial refocus of spin. However, this bump is not observed in the experiment. It is probably due to the fact that in the simulation, we only take into account the gradient in the longitudinal direction. In reality, although gradients in other directions are smaller than the longitudinal one, they still affect the precession frequency of each individual spin. Consequently, spin refocus is disturbed and the small bump is smeared out.

\section{Discussion}

\begin{figure}
		\centering
		\includegraphics[width=8.0cm]{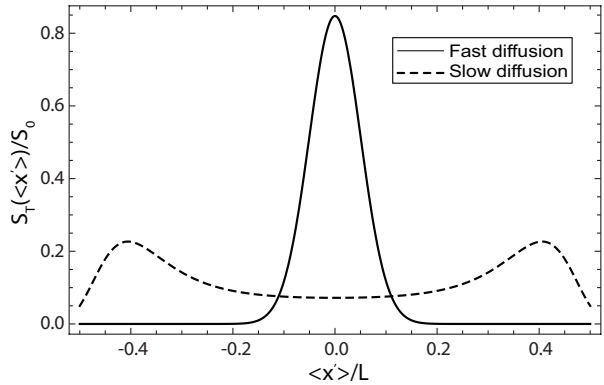}
	\caption{Frequency spectrum of $S_T$ as a function of $\left\langle x^{\prime}\right\rangle/L$ at $t=0.02$ s. The length $L$ of the cell is 1 cm. The solid line corresponds to $D=1000$ cm$^2$/s, which is in the motional averaging regime, and the dashed line corresponds to $D=1$ cm$^2$/s, which is in the slow diffusion limit.}
	\label{Spectrum}
\end{figure}

In Sec. III, we numerically calculate $S_T(\left\langle x^{\prime}\right\rangle)$ without any approximation. The time evolution of integrated $S_T(\left\langle x^{\prime}\right\rangle)$ is compared to the FID signal of polarized $^3$He gas and a good agreement is found, especially in the intermediate regime. It is also interesting to see how $S_T$ changes as a function of position $\left\langle x^{\prime}\right\rangle$ when different diffusion regimes are considered. In Fig. \ref{Spectrum}, we show $S_T$ as a function of $\left\langle x^{\prime}\right\rangle$, which is also equivalent to a frequency spectrum due to the linear relationship between $\omega$ and $\left\langle x^{\prime}\right\rangle$, known as frequency encoding. The two curves shown in Fig. \ref{Spectrum} are calculated at the time instant $t=0.02$ s, with $\gamma G=1000$ rad/s$\cdot$cm, $L=1$ cm and $D=1$ and $1000$ cm$^2$/s, respectively. The $D=1$ cm$^2$/s case is in the slow diffusion limit, and two peaks close to the edges are observed. As diffusion is more restricted at the boundary, the diffusion induced relaxation is suppressed, compared with the relaxation at the center. In contrast, the $D=1000$ cm$^2$/s case is in the fast diffusion limit and only one peak centered at the mean frequency presents, which means most of spins precess at the same frequency and relax at the same rate, i.e. Eq. (\ref{eq:fastT2}). These results show that the approach developed in this manuscript is able to capture all distinct behaviors of the transverse magnetization in different diffusion regimes.

In Sec. II B, it is shown that the longitudinal relaxation rate $1/T_1^G$ differs by a factor of 2 between the slow diffusion limit and the fast diffusion high pressure limit. A possible explanation is that when $t$ is small, i.e. in the slow diffusion limit ($4Dt<<(L/2)^2$), most of the spins do not see walls so spins diffuse freely; when $t$ gets larger, it gets into the fast diffusion limit ($4Dt>>(L/2)^2$), where spins see the wall frequently. As diffusion is more restricted in the fast diffusion limit, the effective diffusion speed is smaller than that in the free diffusion. As a result, the relaxation rate in the fast diffusion limit is smaller, similar to the explanation of the edge enhancement effect. 

\begin{figure}
	\centering
		\includegraphics[width=8.5cm]{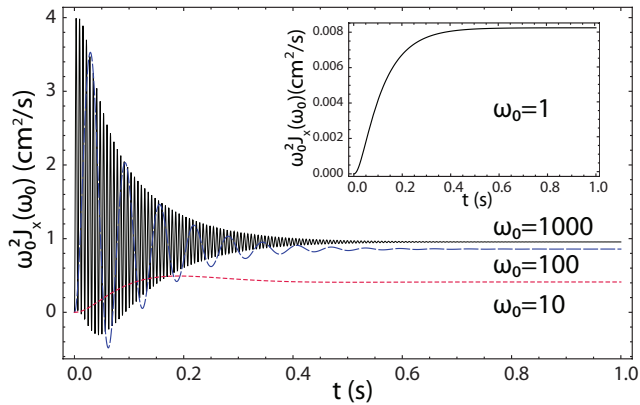}
	\caption{(Color online)The curve with $\omega_0=1000$ rad/s is definitely in the high pressure limit, the normalized relaxation rate evolves into $D$ ($D=1$ cm$^2$/sec) as expected. The inset figure shows the relaxation rate when $\omega_0=1$ rad/s, which is in the low pressure limit. It evolves into $\omega_0^2L^4/120D$, which is 1/120 cm$^2$/s, when $t$ becomes large. The other two curves are in the intermediate region.}
	\label{Jxw}
\end{figure}

We numerically evaluate $\Re[J_x(\omega_0)]$ as a function of time to reveal how $1/T_1^G$ changes from the slow diffusion limit to the fast diffusion limit (Fig. \ref{Jxw}). Values of the parameters used in the evaluation are assigned as $\left\langle x^{\prime}\right\rangle=0$, $D=1$ cm$^2$/s, $L=1$ cm and $\omega_0=1$ to 1000 rad/s. In the figure, the quantity $\omega_0^2\Re[J_x(\omega_0)]$ is actually plotted for the purpose of comparison. Therefore, in the slow diffusion limit, $\omega_0^2\Re[J_x(\omega_0)]=2D$; whereas, in the fast diffusion high pressure limit, $\omega_0^2\Re[J_x(\omega_0)]=D$. As shown in the figure, when $t$ is small, i.e. in the slow diffusion limit, the relaxation rate oscillates around $2D$. When $t$ becomes larger, the oscillating amplitude of the relaxation rate decreases and the mean of the oscillation converges to the fast diffusion results. The final value of the fast diffusion result depends on the ratio of $\tau_d/\tau_l$, see Sec. II B. When $\omega_0=1000$, it is in the high pressure limit and $\omega_0^2J_x(\omega_0)$ converges to $D$, which is 1 cm$^2$/s in our case; and when $\omega_0=1$ rad/s, it is in the low pressure limit and $\omega_0^2J_x(\omega_0)=\omega_0^2L^4/120D=1/120$ cm$^2$/s, see Eq. (\ref{eq:T1eqnlp}). The characteristic time to distinguish the slow diffusion limit from the fast diffusion limit is also $\tau_c=(L/2)^2/4D=0.0625$ s. As $\tau_c$ is usually small in practice, $T_1^G$s measured by experiments are usually in the fast diffusion limit. Nevertheless, When $D$ is small enough or alternatively the cell dimension is large enough, the characteristic time $\tau_c$ can be rather large and it is possible to measure the longitudinal relaxation rate in the slow diffusion regime.

\section{Conclusions}

A new approach based on Redfield theory is developed to calculate magnetic field gradient-induced longitudinal and transverse relaxations of $^3$He gas. As an extension to the method developed by McGregor, the newly developed approach works in all diffusion regimes, including the intermediate regime. It can also explain the edge enhancement effect in the slow and intermediate diffusion regime, which shows the ability to capture	all the relaxation related behaviors in one single unified model. It also has an advantage in terms of numerical simulations, because one only needs to re-compute the probability density function for new geometries. Since the density function is easy to obtain, the relaxation rates can be computed readily as described in this manuscript. 

\section{Acknowledgment}
This work was supported by the School of Arts and Science of Duke University and the U.S. Department of Energy under Contract No. DE-FG02-03ER41231. J.-G. Liu also wants to acknowledge the support of NSF grant DMS-0811177.

\appendix{}
\section{Simplification of the Position Autocorrelation Function in the Slow Diffusion Limit}

In the slow diffusion limit, $4Dt<<(L/2)^2$, Eq. (\ref{eq:xautogensolution}) is simplified by expanding all exponential terms in terms of $t$, $t_0$ and $\tau$ to the first order. Trigonometric functions, such as $\sin(\frac{n\pi \left\langle x^{\prime}\right\rangle}{L})$, can be expanded in terms of $\frac{\left\langle x^{\prime}\right\rangle}{L}$ to the first order. The first term of Eq. (\ref{eq:xautogensolution}) becomes
\begin{equation}
	\frac{8L^2}{\pi^4}\sum^{\infty}_{n=1,3}{\frac{1}{n^4}(1-\frac{n^2\pi^2D\tau}{L^2})}=\frac{L^2}{12}-D\tau.
	\label{eq:term1}
\end{equation}

The second term $\left\langle x^{\prime}\right\rangle^2$ is unchanged, and the third term becomes
\begin{align}
	&\frac{4 \left\langle x^{\prime}\right\rangle L}{\pi^2}\sum^{\infty}_{n=1,3}{\frac{1}{n^2}\sin(\frac{n\pi}{2})\frac{n\pi \left\langle x^{\prime}\right\rangle}{L}(1-\frac{n^2\pi^2Dt}{L^2})}\notag \\
	&=\frac{4 \left\langle x^{\prime}\right\rangle^2 }{\pi}\sum^{\infty}_{n=1,3}{\frac{\sin(\frac{n\pi}{2})}{n}}+O(1/L)= \left\langle x^{\prime}\right\rangle^2.
\end{align}

The fourth term is the same as the third term, which is also $\left\langle x^{\prime}\right\rangle^2$; and the last term becomes
\begin{align}
	\frac{16L^2}{\pi^4}&\sum^{\infty}_{n=1,3}\sum^{\infty}_{k=2,4}{\frac{k^2+n^2}{n^2(k^2-n^2)^2}}\cos(\frac{k\pi}{2})\notag \\
	&\times(1-\frac{n^2\pi^2D\tau}{L^2}-\frac{k^2\pi^2Dt_0}{L^2}-\frac{k^2\pi^2 \left\langle x^{\prime}\right\rangle^2}{2L^2}).
	\label{eq:term5}
\end{align}
The first term in the bracket of Eq. (\ref{eq:term5}) is evaluated to converge to $-L^2/12$. The evaluation of the second term yields $D\tau$. The third term and fourth term are the same, except for different prefactors. They are evaluated to be $2Dt_0$ and $\left\langle x^{\prime}\right\rangle^2$, respectively. Collecting all these terms together, the autocorrelation function of $x$ becomes
\begin{align}	
&\overline{(x(t-\tau)- \left\langle x^{\prime}\right\rangle)(x(t)-\left\langle x^{\prime}\right\rangle)}\notag \\
&=\underbrace{\frac{L^2}{12}-D\tau}_{1st~term}\underbrace{+\left\langle x^{\prime}\right\rangle^2}_{2nd~term}\underbrace{-\left\langle x^{\prime}\right\rangle^2-\left\langle x^{\prime}\right\rangle^2}_{3rd~and~4th~term}\notag \\
&\underbrace{-\frac{L^2}{12}+D\tau+2Dt_0+\left\langle x^{\prime}\right\rangle^2}_{5th~term}+O(1/L)+O(t^2)\notag \\
&=2D(t-\tau). 
\label{eq:A4}
\end{align}

An alternative way to obtain Eq. (\ref{eq:A4}) is to solve the diffusion equation in free space as slow diffusion is equivalent to free diffusion. In this case, the conditional probability function $\rho(x,t|x_0,t_0)$ is known to be
\begin{equation}
	\rho(x,t|x_0,t_0)=\frac{1}{\sqrt{4\pi D\tau}}e^{-\frac{(x-x_0)^2}{4D\tau}}.
	\label{eq:freediff}
\end{equation}
In the free diffusion, the diffusion equation as well as the autocorrelation is translational invariant. Therefore, $\overline{(x(t-\tau)-\left\langle x^{\prime}\right\rangle)(x(t)-\left\langle x^{\prime}\right\rangle)}=\overline{x(t-\tau)x(t)}$ and
\begin{align}
	\overline{x(t-\tau)x(t)}&=\int^{\infty}_{-\infty}dx_{0}\frac{x_{0}e^{-\frac{x^{2}_{0}}{4Dt_{0}}}}{\sqrt{4\pi Dt_{0}}}\int^{\infty}_{-\infty}\frac{xe^{-\frac{(x-x_{0})^{2}}{4D\tau}}}{\sqrt{4\pi D\tau}}dx \notag \\
	&=2D(t-\tau),
\end{align}
which is exactly the same as Eq. (\ref{eq:A4}).

\end{document}